\documentclass{article}

\usepackage[english]{babel}

\usepackage[letterpaper,top=2cm,bottom=2cm,left=3cm,right=3cm,marginparwidth=1.75cm]{geometry}

\usepackage[utf8]{inputenc}
\usepackage{amsmath}
\usepackage{mathtools}
\usepackage{mdframed}
\usepackage{graphicx}
\usepackage[colorlinks=true, allcolors=blue]{hyperref}
\usepackage{paralist}
\usepackage{mathrsfs}
\usepackage[scr=rsfs,cal=boondox]{mathalfa}
\usepackage{bm,nicefrac,xfrac}
\usepackage{multicol}                                          
\usepackage{cite}
\usepackage{authblk}
\usepackage{physics}
\usepackage{comment}  % PW added
\usepackage{leftindex}
\usepackage{stackengine} %
\usepackage{amssymb, rotating}
\usepackage[center]{titlesec}
\usepackage{caption}
\usepackage{subcaption}

\addto{\captionsenglish}{}

\begin{document}

\title{\bf{Status of Birkhoff's theorem in polymerized semiclassical regime of Loop Quantum Gravity}}

\author[1]{Luca Cafaro\thanks{lcafaro@fuw.edu.pl}}
\author[2]{Jerzy Lewandowski\thanks{jerzy.lewandowski@fuw.edu.pl}}
\affil[1,2]{Faculty of Physics, University of Warsaw, Pasteura 5, 02-093 Warsaw, Poland\vspace{-2em} }

\date{} % Leave empty to omit a date

\maketitle

\begin{abstract}
The collapse of a spherically symmetric ball of dust has been intensively studied in Loop Quantum Gravity (LQG). From a quantum theory, it is possible to recover a semiclassical regime through a polymerization procedure. In this setting, general solutions to the Polymerized Einstein Field Equations (PEFE) will be discussed both for the homogeneous interior and the exterior of the dust cloud. Exterior solutions are particularly interesting since they may lead to a semiclassical version of the Birkhoff's theorem. It is seen that if time independence of the vacuum is imposed, there exists a class of solutions depending on two parameters. Nevertheless, the possibility of more intricate time dependent solutions is not ruled out completely.\\
A second approach to study semiclassical spacetimes is by considering an Oppenheimer-Snyder model. Namely, one glues the portion of spacetime containing homogeneous dust with the vacuum part by matching the extrinsic curvatures. In this way, one gets a metric tensor for the vacuum which can be compared to the one obtained previously.\\
Although these two methods are completely independent from each other, the results we obtained are in perfect agreement. 
\end{abstract}
    
%\bigskip
\maketitle

\section{Introduction}

Quantum gravity models lead to semiclassical corrections to spacetime geometry. Those spacetimes are still described by a metric tensor with Lorentzian signature, but this tensor contains new expressions proportional to positive powers of Planck's constant.  Often (though not always) these new terms smooth out the spacetime singularities predicted by the classical theory. In the following work we restrict ourselves to semiclassical spacetimes predicted by  LQG models.  The first family of semiclassical spacetimes, which has been very well studied and described in the literature, consists of spatially homogeneous isotropic  spacetimes filled with matter. They either describe the collapse of a dust ball or the expansion of the universe. A second family includes spherically symmetric vacuum as well and it will be the kingpin of our paper. From the point of view of Einstein's classical theory, all the spherically symmetric vacuum spacetimes are characterized by Birkhoff's theorem as a one-parameter family of spacetimes given by the Schwarzschild metric tensor. This raises the question of how this theorem generalizes (in possibly modified form) into the semiclassical theory.\\
Semiclassical spherically symmetric vacuum spacetimes are obtained in various ways. The direct method involves finding an exact quantum state in which the quantum observables have minimal quantum uncertainties, followed by analyzing the expectation values of the quantum operators composing the metric tensor. 
Alternatively, it is possible to consider suitably modified Einstein's equation and look for solutions.
A third method relies on the Oppenheimer-Snyder (OS) approach. That is to say, the classical collapsing dust ball is replaced with a semiclassical version, and then the static, spherically symmetric metric tensor outside is computed, determined by the known matching conditions. For this reason we will refer to this method as "matched" OS (mOS).
The last two approaches will be the main focus of this paper.\\
At first we will look at semiclassical Einstein equations. The aim is to find the most general solution describing a collapsing ball of homogeneous dust embedded in a vacuum. Interior and exterior of the "star" will be treated separately.\\
Lastly, said solution could be compared to spacetime metrics obtained by the mOS approach. In contrast to the classical mOS models, their semiclassical counterparts provide three different branches of spherically symmetric vacuum, depending on the parameter values $k=-1,0,1$ (open, flat, and closed universe, respectively). In the case of $k=0$, the obtained family of metric tensors outside the dust ball is parameterized by one parameter, the mass. When $k \neq 0$ a second parameter, related to the initial radius of the ball, appears. Its absolute value is the same in both cases, however it is positive if $k=-1$ and negative if $k=1$.\\
Do these three metric satisfy the semiclassical Einstein's equations? Do these equations admit any other vacuum spherically symmetric solutions? Do all the solutions have static asymptotically flat regions?  Finding answers to these questions is the subject of the current paper.

\section{Classical theory}

As it is common in literature, the classical theory will be described through the Hamiltonian formalism and Ashtekar variables.
The starting point is the most general spherically symmetric spacetime which can be expressed in the following way \cite{Gambini_2014}
\begin{equation} \label{Spherical}
    ds^2=-N^2 d\tau^2 + \frac{(E^{\varphi})^2}{E^x} (dx+N^x d\tau)^2 + E^x d\Omega^2 \, .
\end{equation}
The functions $N=N(\tau,x)$ and $N^x=N^x(\tau,x)$ are respectively the Lapse and the x-component of the Shift vector. The other two variables, $E^x=E^x(\tau, x)$ and $E^{\varphi}=E^{\varphi}(\tau, x)$ are related to the densitised triads $E^a_i = \sqrt{q} e^a_i$ through the relations:
\begin{subequations}
\begin{align}
E^x_1=&E^x \sin{\theta}\\
E^\theta_2=&E^{\varphi} \sin{\theta}\\
E^\varphi_3=&E^{\varphi} \, .
\end{align}
\end{subequations}
The dynamical variables $E^x$ and $E^{\varphi}$ are conjugated to the Ashtekar-Barbero connection $A^i_a=\omega_a^i + \gamma K_a^i$, given that $\omega_a^i$ is the spin connection, $K_a^i=K_a^{\; b} e_b^i$ the extrinsic curvature and $\gamma$ the Barbero-Immirzi parmeter.\\
Renaming the component of the extrinsic curvature as:
\begin{subequations}
    \begin{align}
    \gamma K^1_x=&K_x\\
    \gamma K^2_\theta=&K_{\varphi}\\
    \gamma K^3_\varphi=&K_{\varphi} \sin{\theta} \, ,
    \end{align}
\end{subequations}
leads to two couples of conjugated variables $(K_x,E^x)$ and $(K_{\varphi},E^{\varphi})$.\\
\\
Since the main interest of this paper will be focused on LTB and Schwarzschild-like metrics, the line element in \eqref{Spherical} can be Gauge fixed.\\
Firstly, the dust field is taken as the time parameter $\mathcal{T}=\tau$ (Dust Gauge). It follows that $N=1$ is needed in order to preserve the Gauge choice in time \cite{Husain_2012}. Such fixing implies that the gravitational Hamiltonian constraint becomes the physical Hamiltonian (density) $H_{phys}=H^{(g)}=-4 \pi p_{\mathcal{T}}$, where $p_{\mathcal{T}}$ is the conjugate momentum to the dust field. It follows that the dust density is related to the Hamiltonian by $\rho=\frac{p_{\mathcal{T}}}{\sqrt{E^x} E^\varphi}=-\frac{H_{phys}}{4 \pi \sqrt{E^x} E^\varphi }$ (see \cite{Kelly_2020} for details).\\
Additionally, the Areal Gauge will be imposed too, $E^x=x^2$.\\
For convenience, $E^{\varphi}$ will be rewrtten as 
\begin{equation} \label{Ansatz}
E^{\varphi}=\pm \frac{x}{\sqrt{1+\epsilon(\tau,x)}} \, .
\end{equation}
All in all the metric becomes
\begin{equation} \label{PG}
    ds^2=-d\tau^2 + \frac{1}{1+\epsilon(\tau, x)} (dx+N^x d\tau)^2 + x^2 d\Omega^2 \, .
\end{equation}
The coordinates adopted in \eqref{PG} are known as generalised Painlevé-Gullstrand (PG) coordinates $(\tau, x, \theta, \varphi)$.\\
All those models with $\epsilon \neq 0$ are known in literature as "non-marginally bound" \cite{Bojowald_2009, giesel2023generalized, giesel2023embedding} and will be the main interest of this paper.\\
Now, the dynamic of the system is all encoded in one dynamical couple $(K_{\varphi},E^{\varphi})$, whose Poisson bracket reads $\{ K_{\varphi}(y_1), E^{\varphi}(y_2) \}=\gamma G \delta(y_1-y_2)$. From now on the unit system will be set such that $G=1$.\\
\\
From this point, the LTB model in comoving coordinates $(T,R,\theta, \varphi)$ is recovered by choosing $\tau=T$, $x=\xi(T,R)$ and $N^x=-\partial_T \xi(T,R)$
\begin{equation} \label{LTB}
ds^2= -dT^2 + \frac{[ \partial_R \xi (T,R)]^2}{1+E(R)} dR^2 + \xi(t,R)^2 d\Omega^2 \, ,
\end{equation}
with $\epsilon(\tau, x)=E(R)$ (this justifies the choice in \eqref{Ansatz}).\\
It is useful to mention that the Friedmann-Lemaître-Robertson-Walker (FLRW) metric is a particular case of the LTB model.\\
By choosing $\xi(T,R)=a(T)\chi_k(R)$ and $E(R)=-k\chi_k^2(R)$, one gets to
\begin{equation} \label{FLRW}
ds^2= -dT^2 + a^2 dR^2 + a^2 \chi_k^2 d\Omega^2 \, .
\end{equation}
The function $a (T)$ is the scale factor obeying the Friedmann equation, whereas the function $\chi_k (R)$ is defined as
\begin{equation} \label{rad_fun}
\chi_k(R)= 
\begin{cases}
    \frac{1}{\sqrt{k}} \sin{(\sqrt{k} R)},& \text{if} \quad k=1\\
    R              & \text{if} \quad k=0  \; . \\
    \frac{1}{\sqrt{|k|}} \sinh{(\sqrt{|k|} R)}             & \text{if} \quad k=-1
\end{cases}
\end{equation}
with $R \in [0, \pi]$. Notice that for this function, the following identity holds $(\chi_k^{\prime})^2+k\chi_k^2=1$.\\
\\
On the other hand, the metric in \eqref{PG} describes the Schwarzschild solution as well \cite{Kanai_2011}.
Such a metric is indeed recovered by setting $E^x=(E^{\varphi})^2=x^2$ ($\epsilon=0$), $N=1$ and $N^x=\sqrt{\frac{2 M}{x}}$. Then, changing coordinates such that $dx=dr$ and $d\tau=dt+dr \sqrt{\frac{2M}{r} }\frac{1}{1-\frac{2 M}{r}}$ , the Schwarzschild line element is obtained.\\
It will be shown later that one could actually retrieve any metric of the form
\begin{equation} \label{SC}
ds^2=-f(r)dt^2+\frac{1}{f(r)}dr^2+r^2 d\Omega^2 \, ,
\end{equation}
when semiclassical corrections are considered.
\\

\section{Semiclassical collapse of a dust ball}

The problem to be addressed is the collapse of a finite-size dust ball, characterised by a metric provided by \eqref{LTB}, embedded in a vacuum described by \eqref{SC} using quantum correction coming from LQG.
The whole spacetime can be described by a single coordinate system using the PG coordinates as in \eqref{PG}.\\
\\
The quantum theory of such a system relies on a 1-dimensional graph as it has been described in \cite{Husain_2022, Gambini_2023, Olmedo_2016, Bojowald_2008}. A laconic review follows.\\
The couple $(K_{\varphi},E^{\varphi})$ acts on each vertex $x_j$, whereas $(K_x,E^x)$ acts on each edge. Given the Gauge fixing that has been performed at the beginning, only the variables on $\varphi$ will play a role. In the $E^{\varphi}$ representation, one could define a state $\ket{E^{\varphi}_j}_j$ that spans a Hilbert space associated to the vertex $x_j$, i.e. $\mathcal{H}_j$. The full Hilbert space is the tensor product $\mathcal{H}=\otimes_j \mathcal{H}_j$.\\
The scalar product on each $\mathcal{H}_j$ is provided by $\leftindex[I]_j {\bra{\tilde{E}^{\varphi}_j}\ket{E^{\varphi}_j}}_j= \delta_{\tilde{E}^{\varphi}_j E^{\varphi}_j}$. Additionally, states belonging to two different Hilbert spaces (e.g. $\mathcal{H}_i$ and $\mathcal{H}_j$) are automatically orthogonal. \\
It is natural to associate an operator to the triad component $\hat{E^{\varphi}_j}$, while the extrinsic curvature requires a point holonomy operator $\hat{U}_j=e^{i \bar{\mu}_j \hat{K}_{\varphi, j} }$. The scheme adopted here is the $\bar{\mu}$ scheme of Loop Quantum Cosmology \cite{B_hmer_2007, Chiou_2008}, that is $\bar{\mu}_j=\frac{\sqrt{\Delta}}{x_j}$, with $\Delta$ the smallest non-zero eigenvalue of the area operator. 
The action of those operators is given by
\begin{subequations}
\begin{align}
\hat{E}^{\varphi}_j \ket{E^{\varphi}_j}_j=& E^{\varphi}_j \ket{E^{\varphi}_j}_j\\
\hat{U}_j \ket{E^{\varphi}_j}_j=& \ket{E^{\varphi}_j + \bar{\mu}_j}_j \, .
\end{align}
\end{subequations}
Since the inverse of $E^{\varphi}$ appears in the Hamiltonian, one is forced to introduce corrections from the inverse triad operator.
Operationally, given a state $\ket{E^{\varphi}_j}_j$, the eigenvalue of the operator $\hat{\nicefrac{1}{E^{\varphi}_j}}$ is taken to be zero whenever the eigenvalue of $\hat{E}^\varphi_j$ is zero. In all other cases the eigenvalue of the inverse triad operator is $\nicefrac{1}{E^{\varphi}_j}$.\\
The key point to recover a semiclassical theory is the introduction of the polymerization\\ $K_{\varphi, j} \rightarrow \frac{\hat{U}_j-\hat{U}_j^\dag}{2 i \bar{\mu}_j}=\widehat{\frac{\sin{\bar{\mu}_j K_{\varphi, j}}}{\bar{\mu}_j}}$. 
Notice that in the classical limit $(\bar{\mu}_j \rightarrow 0)$, one has $\frac{\sin{\bar{\mu}_j K_{\varphi, j}}}{\bar{\mu}_j} \rightarrow K_{\varphi, j}$.\\
All in all, the semiclassical Hamiltonian density is recovered by writing its corresponding operators in terms of the classical phase space variables and by reintroducing the continuum limit:
\begin{equation} \label{Hamiltonian}
    H_{phys}=-\frac{1}{2} \left[ \frac{E^{\varphi}}{\gamma^2 \Delta x} \partial_x \left( x^3 \sin^2{\beta} \right) +\frac{x}{E^{\varphi}} + \frac{E^{\varphi}}{x} -2 \partial_x \left( \frac{x^2}{E^{\varphi}}\right) \right] \, .
\end{equation}
The full Hamiltonian is recovered upon integration over x.\\
By taking the Posisson brackets of $E^{\varphi}$ and $K_{\varphi}$ with the new Hamiltonian, one recovers the Polymerized Einstein field equations (PEFE). Along with the PEFE, one can write the expressions for the density and the radial component of the shift. All together they read \cite{Kelly_2020, Kelly_2020_, Husain_2022, PhysRevLett.128.121301}
\begin{subequations} \label{PEFE}
\begin{align}
&\dot{E}^{\varphi} =-\frac{x^2}{\gamma \sqrt{\Delta}} \partial_x \left( \frac{E^{\varphi}}{x} \right) \sin{\beta} \cos{\beta}  \label{Eb} \\
&\dot{K}_{\varphi} =\frac{\gamma x}{2 (E^{\varphi})^2}-\frac{\gamma}{2 x} -\frac{\partial_x \left(x^3 \sin^2{\beta}\right)}{2 \gamma \Delta x} \label{b} \\
&\rho =\frac{1}{8 \pi x E^{\varphi}} \left[ \frac{E^{\varphi}}{\gamma^2 \Delta x} \partial_x \left( x^3 \sin^2{\beta} \right) +\frac{x}{E^{\varphi}} + \frac{E^{\varphi}}{x} -2 \partial_x \left( \frac{x^2}{E^{\varphi}}\right) \right]  \label{rho} \\
&N^x = - \frac{x}{\gamma \sqrt{\Delta}} \sin{\beta} \cos{\beta}   \label{Nx}\, ,
\end{align}
\end{subequations}
with $\beta \coloneqq \frac{\sqrt{\Delta}}{x} K_{\varphi} $ and $\dot{()}$ the derivative with respect to time.\\
Alternatively, one could substitute \eqref{b} with a combination of \eqref{b} and \eqref{rho}. Namely
\begin{equation} \label{beta}
    \dot{\beta}=-4 \pi \gamma \Delta \rho +\frac{\gamma \sqrt{\Delta}}{E^{\varphi}} \left[ \frac{1}{E^{\varphi}} - \frac{1}{x} \partial_x \left(\frac{x^2}{E^{\varphi}}\right)\right] \, .
\end{equation}
As it was pointed out in \cite{giesel2023generalized}, the sign of $N^x$ requires a careful analysis in order to avoid discontinuities of the metric. As it will be clarified later, collapsing solutions need the RHS of \eqref{Nx} to be positive. This implies that the product $\sin{\beta} \cos{\beta}$ has to be negative. On the other hand, the expansion requires $\sin{\beta}$ and $\cos{\beta}$ to be equal in sign. 
As a consequence, at the beginning one can set $\sin{\beta}>0$ and, since the dust is collapsing, $\cos{\beta}<0$ (see figure \ref{fig:beta}).\\
Given the equations for the metric, it is suitable to study separately the interior and the exterior of the dust ball.

\subsection*{Interior}

The interior is distinguished by $\rho \neq 0$. Additionally, we restrict our discussion to the homogeneous sector of the LTB model, also known as Oppenheimer-Snyder, which means $\partial_x \rho=0$.\\
The starting point is to plug the ansatz \eqref{Ansatz} into \eqref{rho} and solve it for $\sin{\beta}$. It results in
\begin{equation} \label{sin_int}
    \sin^2{\beta} = \frac{8 \pi \gamma^2 \Delta}{3} \rho + \frac{\gamma^2 \Delta}{x^2} \epsilon \, .
\end{equation}
Notice that, in this first phase of collapse, $\sin{\beta}$ is imposed to be positive, hence $\cos{\beta}$ is negative.\\
From this result, equation \eqref{Eb} can be morphed into
\begin{equation} \label{eps_int}
    \dot{\epsilon}= \epsilon^{\prime} \sqrt{\frac{8 \pi}{3} \rho x^2 + \epsilon} \sqrt{1-\frac{\rho}{\rho_c}-\frac{3}{8 \pi \rho_c} \frac{\epsilon}{x^2}} \, ,
\end{equation}
where $\rho_c \coloneqq \frac{3}{8 \pi \gamma^2 \Delta}$.\\
\\
In LTB coordinates $(T,R)$, one has $x=\xi(T,R)$ and $\epsilon(\tau, x)=E(R)$. It is easy to verify that $N^x=\frac{\dot{\epsilon}}{\epsilon^{\prime}}=-\partial_T \xi$. Therefore \eqref{eps_int} can be rewritten as 
\begin{equation} \label{LTB_fried}
    \frac{\partial_T \xi}{\xi}=- \sqrt{\frac{8 \pi}{3} \rho + \frac{E}{\xi^2}} \sqrt{1-\frac{\rho}{\rho_c}-\frac{3}{8 \pi \rho_c} \frac{E}{\xi^2}} \, ,
\end{equation}
which is general for any OS model. The "-" sign in the RHS of \eqref{LTB_fried} represents a collapse. This has been made possible by choosing $\sin{\beta}\cos{\beta}<0$ as it was stated before.\\
A simplification occurs by imposing $x=\xi(T,R)=a(T)\chi_k(R)$. Given that, in (T, R) coordinates, the time dependence disappears from $E(R)$, it is expected that $\epsilon(\tau, x)$ is a generic function of the form $\epsilon=h(\frac{x}{a})$. Therefore, equation \eqref{eps_int} becomes
\begin{equation}
    \left( \frac{\dot{a}}{a}\right)= - \sqrt{\frac{8 \pi}{3} \rho + \frac{h}{x^2}} \sqrt{1-\frac{\rho}{\rho_c}-\frac{3}{8 \pi \rho_c} \frac{h}{x^2}} \, .
\end{equation}
The scale factor $a$ is a function solely of the time parameter $\tau=T$, which means that the LHS is independent of x. In order to cancel the x dependence from the RHS, it is needed to impose $h \propto x^2$. The most general form of h is then $h=e_0 \frac{x^2}{a^2}$, with $e_0$ a constant. The collapse is then described by the following Friedmann equation
\begin{equation} \label{FLRW_1}
    \left( \frac{\dot{a}}{a}\right)=- \sqrt{\frac{8 \pi}{3} \rho + \frac{e_0}{a^2}} \sqrt{1-\frac{\rho}{\rho_c}-\frac{3}{8 \pi \rho_c} \frac{e_0}{a^2}} \, .
\end{equation}
In $(T, R)$ coordinates, one has $E(R)=e_0 \chi_k^2(R)$ and the LTB line element is 
\begin{equation} \label{FLRW_2}
ds^2= -dT^2 + \frac{a^2 (\chi_k^{\prime})^2}{1+e_0 \chi_k^2} dR^2 + a^2 \chi_k^2 d\Omega^2 \, .
\end{equation}
Without loss of generality, it is possible to write $e_0$ as $e_0=-k \tilde{e}_0$, with $k=0, \pm 1$ and $\tilde{e}_0 > 0$. Introducing a new variable $\hat{R}= \sqrt{\tilde{e}_0} \chi_k$, one has
\begin{equation} \label{FLRW_3}
ds^2= -dT^2 +\frac{1}{\tilde{e}_0} \frac{a^2}{1-k \hat{R}}d\hat{R}^2 + \frac{1}{\tilde{e}_0} a^2 \hat{R} d\Omega^2 \, .
\end{equation}
Finally, the new constant can be reabsorbed in the scale factor so as to have the FLRW metric in reduced-circumference coordinates.
The line element \eqref{FLRW_2} can be therefore rewritten as the classical one in \eqref{FLRW} provided that $a(T)$ now obeys the modified Friedmann equation \eqref{FLRW_1}.
\\
As a consistency check, one can verify that $\rho \propto a^{-3}$ as it is expected for dust. By differentiating equation \eqref{sin_int} with respect to $\tau$ and comparing it with \eqref{beta}, one is led to 
\begin{equation} \label{rho_1}
    \dot{\rho}=\frac{3}{x}\frac{\dot{\epsilon}}{\epsilon^{\prime}} \rho \, .
\end{equation}
Provided that $\epsilon=e_0 \frac{x^2}{a^2}$, one finds $\rho=\frac{\mathcal{C}}{a^3}$, with $\mathcal{C}=const$.\\

\subsection*{Exterior}

\subsubsection*{Time independent exterior}
\begin{figure} 
    \centering
     \begin{subfigure}[h!]{0.3\textwidth}
         \centering
         \includegraphics[width=1.3\textwidth]{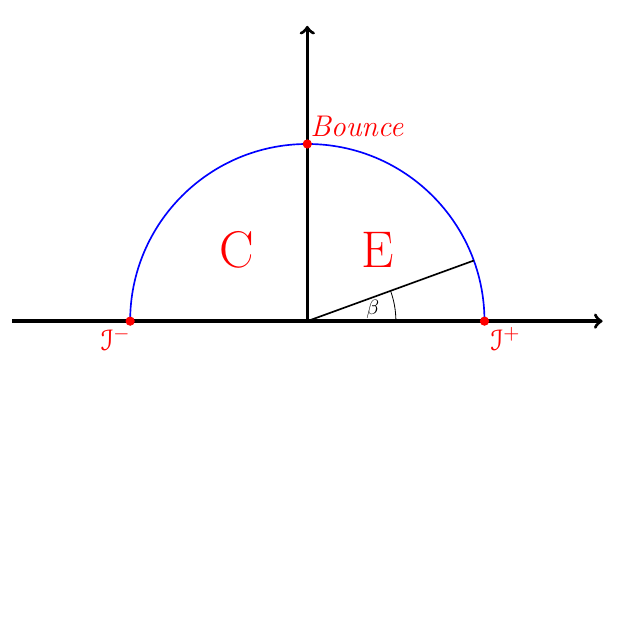}
         %\caption{$y=x$}
         \label{fig:posB}
     \end{subfigure}
     \hspace{3em}%
     \begin{subfigure}[h!]{0.5\textwidth}
         \centering
         \includegraphics[width=0.8\textwidth]{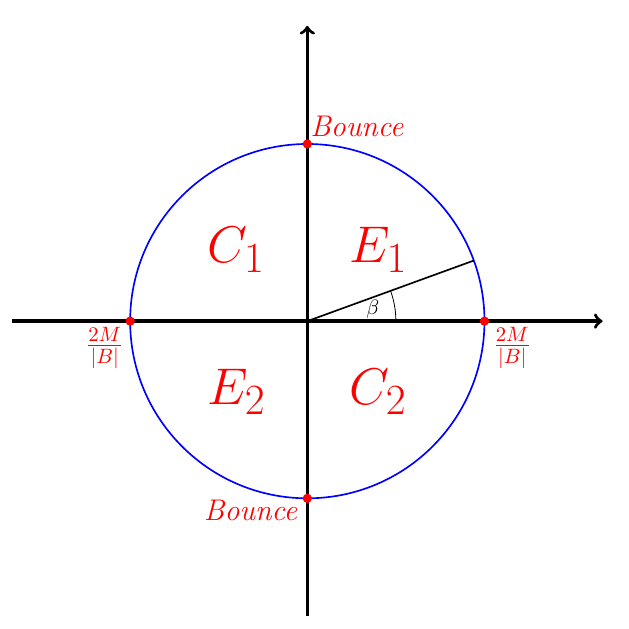}
         %\caption{$y=3\sin x$}
         \label{fig:negB}
     \end{subfigure}
     %\captionsetup{singlelinecheck=false} % <---
     \caption{Display of the different values of $\beta$ throughout the spacetime. The letter "C" stands for "collapse" whereas "E" is for "expansion". \\
     On the left, the case for $B>0$. At the bouncing point $\sin^2{\beta}=1$, therefore $\cos{\beta}=0$. At spatial infinity, $\sin^2{\beta}=0$, then $\cos{\beta}=\pm 1$. Since $\sin{\beta}$ is kept positive, the collapsing solution is provided by $\cos{\beta}<0$. The opposite happens for the expansion in the post-bounce phase. Notice that the very same result could have been achieved by imposing $\sin{\beta}<0$. In this setting $\cos{\beta}<0$ would represent the expansion whereas $\cos{\beta}>0$ the collapse.\\
     If $B<0$, on the right, once again the dust starts in the sector $C_1$ bounces and reaches $E_1$. In this case however the dust reaches a certain value of density (at a certain radius) and recollapses. Consequently, the dust enters the new collapsing region $C_2$, followed by $E_2$ so as to start again from $C_1$.}
     \label{fig:beta} 
\end{figure}

Exterior solutions to the PEFE are found by imposing $\rho=0$.\\
At first, one can look for time independent solutions, i.e. $\dot{E}^{\varphi}=0$.
It follows that it is either $E^{\varphi}=Ax$ ($A = const \neq 0$), $\sin{\beta}=0$ or $\cos{\beta}=0$.
Nevertheless, the second solution implies $E^{\varphi}=x$, e.g. $A=1$. As a matter of fact, $E^{\varphi}=x$ is recovered for $x^3 \sin^2{\beta}=const$, which includes the case $const=0$.\\ 
Finally, the third option $\cos{\beta}$ runs into inconsistencies.\\
For this reasons the only interesting solution is $E^{\varphi}=Ax$.\\
Notice that the latter is recovered by imposing $\epsilon=B=\frac{1}{A^2}-1$ in \eqref{Ansatz}.\\
From inserting this last condition in \eqref{beta}, it is found $\dot{K}_{\varphi}=0$ which is consistent with the time independence of the metric.\\
The exterior analogue of \eqref{sin_int} is 
\begin{equation} \label{sin_ext}
    \sin^2{\beta} = \frac{\gamma^2 \Delta}{x^2} B + \frac{C}{x^3} \, 
\end{equation}
with $C$ an integration constant. Its value is recovered by looking at the Schwarzschild limit ($\Delta \rightarrow 0$), which corresponds to $C=2\gamma^2 \Delta M$.\\
The same value for $C$ can also be recovered from a suitably modified ADM mass. Classically it is defined as
 \cite{Olmedo_2016}:
\begin{equation} \label{mass}
    M= \frac{1}{2} \sqrt{E^x} (1+ \frac{K_{\varphi}^2}{\gamma^2})-\frac{\sqrt{E^x} [(E^x)^{\prime}]^2}{8 (E^{\varphi})^2}\, .
\end{equation}
If one performs the substitution $K_{\varphi}  \rightarrow \frac{\sin{\bar{\mu} K_{\varphi}}}{\bar{\mu}}$ (with $\bar{\mu}=\frac{\sqrt{\Delta}}{x}$),  then a quick calculation shows that $M=\frac{C}{2 \gamma^2 \Delta}$, which is the value shown above.\\
It must be noticed that since $\sin^2{\beta} \ge 0$, one has $B>0$ in order to satisfy the disequality at every $x$. This is equivalent to say that $A^2 < 1$, hence $-1 < A < 1$.\\
On the other hand, it is still possible to pick $B<0$ but restrict the solutions at $x$ lower than a certain $x_{max}$.\\
The complementary bound, $\sin^2{\beta} \le 1$ sets the existence of a minimal $x$,  $x_b=\left( 2\gamma^2 \Delta M\right)^{\nicefrac{1}{3}}+ \frac{B}{6M}\left( 2\gamma^2 \Delta M\right)^{\nicefrac{2}{3}} + O(\Delta^{\nicefrac{4}{3}})$. This value of the radius will coincide with the bouncing radius studied in the last section.\\
Substituting \eqref{sin_ext} into \eqref{Nx} leads to
\begin{equation}
    \left( N^x\right)^2=\frac{2M}{x}-\frac{\alpha}{x^2}\left( \frac{M}{x} + \frac{B}{2}\right)^2 + B \, ,
\end{equation}
where $\alpha$ has been defined as $\alpha=4 \gamma^2 \Delta$.\\
This last equation determines the metric completely. However, it is more convenient to change coordinate so as to have a line element like \eqref{SC}. This is obtained by setting $dx=dr$ and $d\tau=\frac{1}{A}dt + \frac{N^x}{f(r)}dr$, with 
\begin{equation} \label{SC-like}
    f(r)=\frac{1}{A^2}-(N^x)^2=1-\frac{2M}{r}+\frac{\alpha}{r^2}\left( \frac{M}{r} + \frac{B}{2}\right)^2 \, .
\end{equation}
\\
In the case where $A= \pm 1$, then $B=0$.\\
The conclusion is that the family of solutions to the equations \eqref{PEFE} (the semiclassical modification of Einstein's equations), such that $\epsilon = {\rm const}$, is mapped by a coordinate transformation into the family of metric tensors \eqref{SC} defined by \eqref{SC-like} for all the values of the constants $M$ and $B$. A catch is that, according to equation \eqref{sin_ext}, the radial coordinate $x=r$ is to be restricted either to $r_b \le r$ for $B\ge 0$ or $r_b\le r\le \frac{2M}{|B|}$ for $B<0$. However, the resulting metric tensor \eqref{SC-like} is analytic and there is no obstruction for considering it for the other  values of $r$ as long as $f$ is finite, or extending the spacetime even more by using the Eddington-Finkelstein coordinates.\\
One last remark on the sign of $\sin{\beta}$ and $\cos{\beta}$ is in order.
As it was mentioned at the beginning of this section, as the collapse starts, the $\sin{\beta}$ is kept positive while $\cos{\beta}$ is negative in the collapsing phase and positive during the expansion.
However, what happens next is dictated by the sign of $B$. In the last section of this paper, it will be discussed that for $B>0$ the matter bounces and reaches spatial infinity. On the other hand, a negative $B$ forces the matter to undergo an infinite series of bounces (see the discussion of $k=1$ for the Oppenheimer-Snyder model in the last section). This is portrayed in figure \ref{fig:beta}.\\
\\

\newpage

\subsubsection*{Time dependent exterior}

A more general solution for the exterior is found by restoring the time dependence of $E^{\varphi}$. We can impose \eqref{Ansatz} with general $\epsilon$, as for the interior, and follow the same procedure.
One finds:
\begin{align} 
    &\sin^2{\beta} = \frac{\gamma^2 \Delta}{x^2} \epsilon + \frac{2 \gamma^2 \Delta M}{x^3} \label{sin_ext_time} \\
    &\dot{\epsilon} = \epsilon^{\prime} \sqrt{\epsilon + \frac{2 M}{x}} \sqrt{1-\gamma^2 \Delta \left( \frac{\epsilon}{x^2}+\frac{2 M}{x^3} \right)}  \label{eps_ext} \, .
\end{align}
Another useful equation follows from \eqref{beta}:
\begin{equation} \label{dot_beta_ext}
    \dot{\beta}=-\frac{\gamma \sqrt{\Delta}}{2 x} \epsilon^{\prime} \, .
\end{equation}
\\
At first, one could check singularly the cases $\epsilon=\epsilon(x)$ and $\epsilon=\epsilon(\tau)$.\\
\begin{itemize}
  \item $\epsilon=\epsilon(x)$:\\
  From equation \eqref{sin_ext_time}, it appears that $\beta$ too is a function solely of x, implying $\dot{\beta}=0$. However, this is in contrast with equation \eqref{dot_beta_ext}.\\
  Therefore, $\epsilon=\epsilon(x)$ is ruled out and the square roots appearing in \eqref{eps_ext} are non-zero. It follows that the same formula can be inverted to have an expression for $\epsilon^{\prime}$.
  \item $\epsilon=\epsilon(\tau)$:\\
  Since $\epsilon^{\prime}=0$, it is straightforward to see, from equation \eqref{eps_ext}, that $\dot{\epsilon}=0$. Therefore, this possibility too can be ruled out.
\end{itemize}
From this very short analysis, it is to be concluded that either $\epsilon=const$ (retrieving the time independent case discussed above), or $\epsilon=\epsilon(\tau, x)$.\\
However, the time dependence of the metric components does not exclude the existence of a timelike Killing vector field.

\section{PEFE and matched Oppenheimer-Snyder model}

The goal of this section is to glue together the cloud of collapsing semiclassical dust and a spherically symmetric, static exterior of the form \eqref{SC}. In order to do so, it is possible to match the intrinsic metrics and extrinsic curvatures at the surface of the ball. This will determine uniquely the time/radial component of the exterior metric, $f(r)$. That result was also obtained in  \cite{Lewandowski_2023}. Next, we will compare the derived function $f$  with the exterior solutions  \eqref{SC-like}. \\
The semiclassical dust ball is described by \eqref{FLRW} obeying the two Friedmann equations
\begin{subequations} 
\begin{align}
    \left( \frac{\dot{a}}{a}\right)^2 =&\left( \frac{8 \pi}{3} \rho - \frac{k}{a^2} \right)\left(1-\frac{\rho}{\rho_c}+\frac{3}{8 \pi \rho_c} \frac{k}{a^2} \right) \label{F1} \\
    \begin{split} 
    \left( \frac{\ddot{a}}{a}\right) \; =& 
    -\frac{4 \pi}{3} \rho \left(1-\frac{\rho}{\rho_c}+\frac{3}{8 \pi \rho_c} \frac{k}{a^2} \right) + \\
    &+\left( \frac{8 \pi}{3} \rho - \frac{k}{a^2} \right)\left(\frac{3}{2}\frac{\rho}{\rho_c}-\frac{3}{8 \pi \rho_c} \frac{k}{a^2} \right)
    \label{F2} \, ,    
    \end{split}
\end{align}
\end{subequations}
where the second one is the derivative of the first. The density $\rho$ for dust reads $\rho=\frac{\mathcal{C}}{a^3}$.\\
As mentioned at the beginning, this metric is junctioned to an exterior of the kind \eqref{SC}, by applying the Israel condition so as to match the two extrinsic curvatures \cite{geller2018mass, Kwidzinski_2020}. In the end, one finds
\begin{equation}  \label{FLRW_ext}
    f(r)=1-\frac{2M}{r}+\frac{\alpha}{r^2}\left( \frac{M}{r} - \frac{k \chi^2_{k,0}}{2}\right)^2 \, ,
\end{equation}
where $\chi_{k,0}$ is the radial function $\chi_k(R)$ \eqref{rad_fun} at the surface of the ball, e.g. initial radius. From now on the subscript "0" will be used to label the variables evaluated on the surface of the ball.\\
It is easy to verify that this last function is exactly the one recovered from the PEFE \eqref{SC-like} provided that $B=-k\chi^2_{k,0}$.
Notice that  we do not require in this section, that the exterior metric tensor satisfies any equations. The function \eqref{FLRW_ext}  is determined by the junction conditions and by an  assumption that the vector field $\partial_t$ is a Killing vector. This is a direct generalisation of the model studied in \cite{Ziaie_2020, Parvizi:2021ekr, Lewandowski_2023}.\\
The junction of the two line elements leads to $M=\frac{4 \pi}{3}\rho_0 r^3_0 = \frac{4 \pi}{3} \mathcal{C} \chi_{k,0}^3$.\\
The horizons of such a metric are provided by real zeros of \eqref{FLRW_ext}. In the end one has
\begin{subequations} 
    \begin{align}
    r_- &=\left( \frac{\alpha M}{2} \right)^{\nicefrac{1}{3}} + \frac{1-2k\chi_{k,0}^2}{6M} \left( \frac{\alpha M}{2} \right)^{\nicefrac{2}{3}} + \frac{ (1-k\chi_{k,0}^2)^2}{24 M} \alpha + O(\alpha^{\nicefrac{4}{3}}) \label{r-} \\    
    r_+&=2M-\frac{(1-k\chi_{k,0}^2)^2}{8 M}\alpha + O(\alpha^{\nicefrac{4}{3}})
    \label{r+} \, .   
    \end{align}
\end{subequations}
The exact expression of $r_+$ contains a square root whose positivity of the argument is assured when $M \le M_-$ or $M \ge M_+$. The lower and upper bound of $M$ are functions of $k \chi_{k,0}^2$ which read
\begin{equation}  \label{masses}
    M_\pm = \frac{\sqrt{\alpha}}{6 \sqrt{6}} \sqrt{64 -96 k \chi_{k,0}^2+ 30 k^2 \chi_{k,0}^4 +k^3 \chi_{k,0}^6 \pm \left(16-16 k \chi_{k,0}^2 +k^2 \chi_{k,0}^4 \right)^{\nicefrac{3}{2}}} \, .
\end{equation}
It is easy to verify that if $k=0$, then $M_-=0$ and $M_+=\frac{4}{3\sqrt{3}}\sqrt{\alpha}$ (consistent with the result found in \cite{Lewandowski_2023}). Their behaviour for $k=\pm 1$ will be discussed separately below.\\ 
\\ 
The bouncing radius too can be extracted from \eqref{FLRW_ext} by using the geodesic equation. For a massive dust particle, one has $\dot{r}^2=1-k\chi_{k,0}^2-f(r)$. The turning point occurs when $\dot{r}=0$, i.e. $r_b=\left( \frac{\alpha M}{2}\right)^{\nicefrac{1}{3}}-\frac{k \chi_{k,0}^2}{6 M}\left( \frac{\alpha M}{2} \right)^{\nicefrac{2}{3}} + O(\Delta^{\nicefrac{4}{3}})$, same value found using the PEFE with $B=-k \chi^2_{k,0}$. \\
This is not surprising as one can show that $\dot{r}^2=1-k\chi_{k,0}^2-f(r)=0$ is equivalent to $\sin^2{\beta}=1$ \eqref{sin_ext} with $B=-k\chi_{k,0}^2$.\\ The first equation leads to $\dot{r}^2=(-k\chi_{k,0}^2+\frac{2M}{r})-\frac{\alpha}{4 r^2}\left(\frac{2M}{r}-k\chi_{k,0}^2\right)^2=0$. Dividing by $\left(\frac{2M}{r}-k\chi_{k,0}^2\right)$ (we assume here that it is $\neq 0$) brings us to $\frac{\alpha}{4 r^2}\left(\frac{2M}{r}-k\chi_{k,0}^2\right)=1$, which is exactly \eqref{sin_ext}. 
\\
\subsection*{General properties of the matched Oppenheimer-Snyder model with $k \neq 0$} 

As it has already been discussed, the interior metric \eqref{FLRW} is an exact solution to the PEFE with $\epsilon(\tau,x)=E(R)=- k \chi_k^2(R)$.\\
The exterior \eqref{FLRW_ext} instead requires a more detailed analysis. Hence the cases of $k= \pm 1$ will be studied separately ($k=0$ can be found in \cite{Lewandowski_2023}).\\

\subsubsection*{k=1}
\begin{figure}[ht]
\centering
\includegraphics[width=0.7\textwidth]{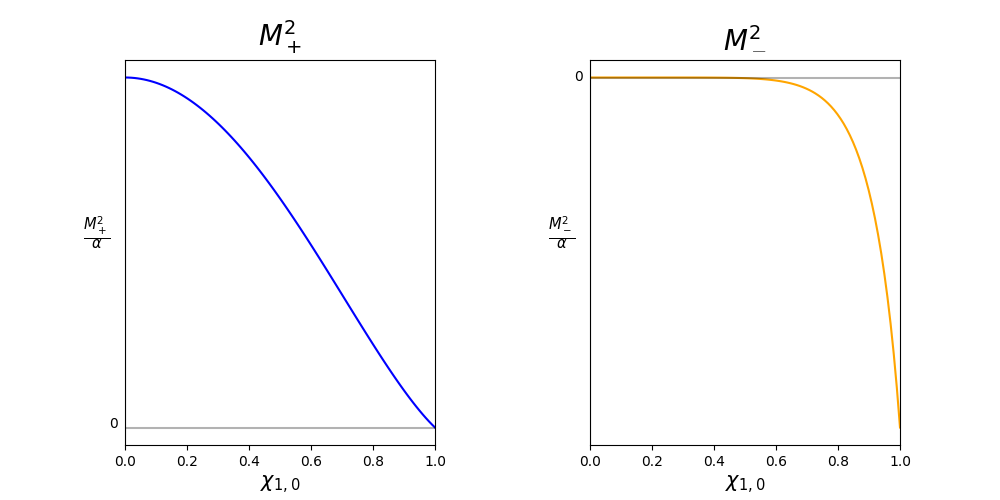}
\caption{\label{fig:kp} Profiles of $M_+^2$ and $M_-^2$ as functions of $\chi_{1,0}$. If $k=1$, $\chi_{1,0}$ is a $\sin$ and therefore goes from 0 to 1  ($R \in [0, \pi]$).}
\end{figure}

Let the ball be motionless at $T=T_0$, implying $a_0=a(T=T_0)=\frac{8 \pi}{3} \mathcal{C}$ (value for which the RHS of \eqref{F1} vanishes). It can be checked that $\ddot{a}<0$, so the ball starts its collapse. Let $M$ be the mass of the ball and $\chi_{1,0}$ its initial radius. Finally, let $\rho_0 \ll \rho_c$.\\
The complete set of initial condition is then
\begin{subequations} \label{initial}
\begin{align}
a_0=&\frac{2M}{\chi^3_{1,0}} \, , \quad \dot{a}_0=0 \, , \quad \ddot{a}_0<0 \\
r_0=&a_0 \chi_{1,0} = \frac{2M}{\chi^2_{1,0}}\\
\rho_0=&\frac{3}{32 \pi}\frac{\chi^6_{1,0}}{M^2} \, .
\end{align}
\end{subequations}
Notice that the constant $\mathcal{C}$ has been replaced by $M$ through the relation given at the beginning.\\
\\
The exterior solution is of the form \eqref{SC-like} but the constant $B=- \chi_{1,0}^2$ is now negative. Nevertheless, it still satisfies the PEFE as long as $\sin^2{\beta} =\gamma^2 \Delta \left( -\frac{k \chi_{1,0}^2}{r^2}  + \frac{2 M}{r^3} \right) \ge 0$ \eqref{sin_ext}. This leads to $r \le r_{max}=\frac{2M}{\chi^2_{k,0}}$, which is exactly $r_0$.\\
In other words, at the starting point $T=T_0$, the ball of dust occupies all the region contained in $r_0$ and it satisfies the interior PEFE (iPEFE). At a subsequent time, the ball occupies a spherical subregion of radius $r_1 < r_0$. 
It implies that the metric restrained by $r_1$ is solution of the iPEFE whereas the ePEFE are satisfied between $r_1$ and $r_0$.\\
The metric outside $r_0$ is not solution to the Einstein equations.\\
\\
From equations \eqref{F1} and \eqref{F2}, one can study the evolution of the system. It turns out that when $\rho$ approaches $\rho_c$ (i.e. when $a$ decreases),  $\ddot{a}$ changes sign and the ball re-expand as soon as the other zero of \eqref{F1} is touched, $\rho=\rho_c + \frac{3}{8 \pi } \frac{1}{a^2_b}$. The value of $a_b$ is recovered by solving the third order equation obtained by replacing $\rho \propto a^{-3}$.\\
The expansion of the dust cloud will stop when reaching once again $a=a_0$, where $\dot{a}=0$ and $\ddot{a}<0$. At this point it will recollapse as it did at the beginning \cite{Corichi_2011}.\\
\\
Horizons are formed when $M<M_-$ and $M>M_+$. From their expressions, one finds the profile in figure \ref{fig:kp}.\\
\\
\\
Given that $M_-^2<0$, the only condition to have a horizon is $M>M_+$. Surprisingly, the actual value of the critical mass is dictated solely by its initial radius and $M_+=0$ if $\chi_{k,0}=1$. On the other hand, if $\chi_{k,0}=0$ the critical mass is the same as in $k=0$. This is due to the fact that $M_+$ is a function of the product $k \chi_{k,0}^2$. 

\subsubsection*{k=-1}
\begin{figure}[ht]
\centering
\includegraphics[width=0.7\textwidth]{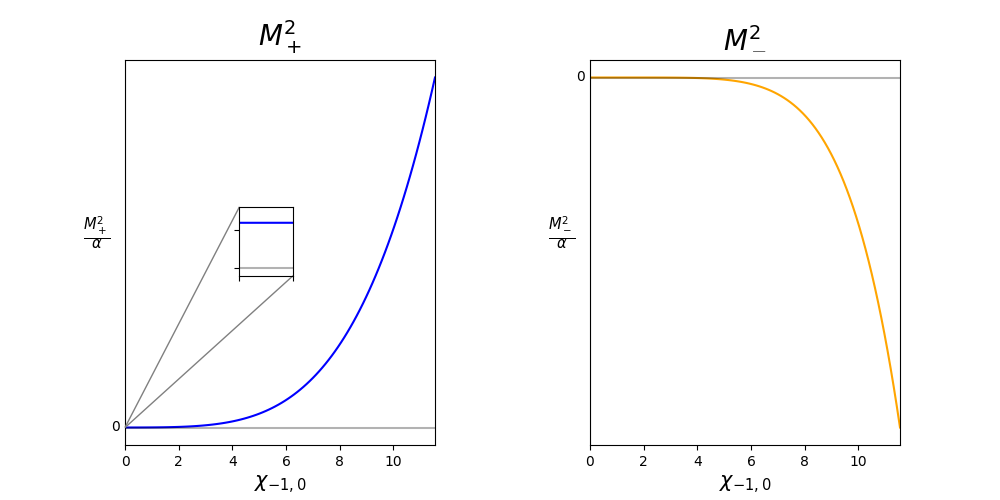}
\caption{\label{fig:km} Profiles of $M_+^2$ and $M_-^2$ as functions of $\chi_{-1,0}$. This time, $\chi_{-1,0}$ is a $\sinh$ and therefore goes from 0 to $\infty$.}
\end{figure}

When $k$ is negative, the RHS of \eqref{F1} cannot vanish for any value of $a$ such that $\rho \ll \rho_c$ and there exists only one zero of $\dot{a}$ (as in $k=0$). The ball will theeqrefore contract for a while but eventually it is bound to re-expand to infinity.\\
Unlike the $k=1$ case though. This is everywhere a solution to the PEFE (both interior and exterior).\\
\\
For what concerns the critical mass, this scenario is pretty similar to the previous one for $M_-$. Once again $M_-^2<0$ for every $\chi_{-1,0}$, so a horizon can form if $M>M_+$. Nevertheless, this time there is no value $\chi_{k,0}$ for which the critical mass is zero (see figure \ref{fig:km}).\\

\section{Discussion}

The semiclassical modification of Einstein's equations has been studied both in the presence of homogeneous dust (interior) and in the vacuum (exterior).\\
In the first case, one gets an OS model whose collapse is dictated by a generalised Friedmann equation \cite{giesel2023generalized, fazzini2023shellcrossings}. This scenario predicts a bounce once the density of dust reaches a certain value. A further generalization to this model occurs when inhomogeneity of the dust is imposed \cite{fazzini2023shellcrossings}.\\
On the other hand, when no dust is present, assuming that the function $\epsilon$ (\ref{Ansatz}) is constant,  we determined  a two-parameter class of static solutions.  Such metrics are dependent on two free parameters, $M$ and $B$. These line elements are  suitably modified Schwarzschild solutions which coincide with Schwarzschild in the classical limit ($\Delta \rightarrow 0$).
\\
Nevertheless, if the function $\epsilon$ is not constant, it is subject to the partial differential equation in \eqref{eps_ext}. It has been discussed that solutions to said equation are either constant or depending on both time and radial coordinate. If the latter case can be (somehow) ruled out, then there exists a unique solution which is time independent and Schwarzschild-like. \\
The static exterior metric is clearly asymptotically flat. It is conceivable, that if $\epsilon$ is not a constant, the situation might change drastically. We do not exclude the existence of a temporal Killing field in this case as well. Nonetheless, it must be taken into account that all of these results strongly depend on the choice of quantization \cite{Singh_2013, Corichi_2011}, the polymerization scheme \cite{Zhang:2023yps} and inverse triad corrections in the quantum theory. 
Different approaches rooted in polymerization techniques are also possible (see for example \cite{Barca_2024}).\\
\\
This analysis can be compared to a model of spherically symmetric collapse obtained in a completely different way. It is indeed possible to match a cloud of homogeneous dust (described by a Friedmann metric) to a general line element of the form \eqref{SC}.\\
It turns out that this model is an exact solution to the PEFE when $\rho \neq 0$. In the vacuum instead, it satisfies globally the PEFE when $k=0, -1$ and just locally when $k=1$.\\
The latter may suggest that the generalized PG coordinates simply fail if $k=1$. This assumption is backed up by the fact that in Schwarzschild-AdS metric the situation is quite similar. The non diagonal piece of the metric in PG coordinates reads $\sqrt{\frac{2M}{r}+\frac{\Lambda r^2}{3}}$ (with $\Lambda <0$). As a consequence, the metric is not well defined if $r$ big enough.\\
On the other hand, it must not be excluded that by correcting the quantum theory, one can extend the $k=1$ case to a global solution without changing coordinates.\\

\section{Acknowledgements} Authors would like to thank Ed Wilson-Ewing and Cong Zhang for numerous discussions. The research presented here was funded by the National Science Centre, Poland as part of the OPUS 24 grant  number 2022/47/B/ST2/02735.

\bibliography{main}
\bibliographystyle{unsrt}

\end{document}